\documentclass[a4paper,fleqn]{cas-dc}

\usepackage[numbers]{natbib}
\usepackage[switch]{lineno}
\newcommand{\absdiv}[1]{%
  \par\addvspace{.5\baselineskip}
  \noindent\textit{#1}\quad\ignorespaces
}
\usepackage{tabularx}
\usepackage{amsmath}
\usepackage{soul,color}
\soulregister\cite7
\usepackage{graphicx}
\usepackage{dcolumn}
\usepackage{array}
\usepackage{algorithm} 
\usepackage{algpseudocode} 
\usepackage{mathtools}
\usepackage{siunitx}
\usepackage{hyphenat}
\usepackage{bm}

\begin{document}

\let\WriteBookmarks\relax
\def\floatpagepagefraction{1}
\def\textpagefraction{.001}
\shorttitle{Universal description of wetting on multiscale surfaces using integral geometry}
\shortauthors{C. Sun et~al.}

\title [mode = title]{Universal description of wetting on multiscale surfaces using integral geometry}   

\author[1]{Chenhao Sun}
\author[2]{James McClure}
\author[3,4]{Steffen Berg}
\author[5]{Peyman Mostaghimi}
\author[5]{Ryan T. Armstrong}
\cormark[1]
\ead{E-mail address: ryan.armstrong@unsw.edu.au}

\address[1]{\normalsize \itshape State Key Laboratory of Petroleum Resources and Prospecting, China University of Petroleum, Beijing, China, 102249}

\address[2]{\normalsize \itshape Advanced Research Computing, Virginia Tech, Wright House, W. Campus Drive, Blacksburg, Virginia, USA, 24061}

\address[3]{\normalsize \itshape Shell Global Solutions International B.V., Grasweg 31, 1031 WG Amsterdam, Netherlands}

\address[4]{\normalsize \itshape Imperial College London, Department of Earth Science \& Engineering and Chemical Engineering, Exhibition Rd, South Kensington, London SW7 2BX, United Kingdom}

\address[5]{\normalsize \itshape School of Minerals \& Energy Resources Engineering, University of New South Wales, Kensington, New South Wales 2052, Australia}
\begin{abstract}
\absdiv{\it Hypothesis}
\newline
Emerging energy-related technologies deal with multiscale hierarchical structures, intricate surface morphology, non-axisymmetric interfaces, and complex contact lines where wetting is difficult to quantify with classical methods. We hypothesis that a universal description of wetting on multiscale surfaces can be developed by using integral geometry coupled to thermodynamic laws. The proposed approach separates the different hierarchy levels of physical description from the thermodynamic description, allowing for a universal description of wetting on multiscale surfaces.
\absdiv{\it Theory and Simulations}
\newline
The theoretical framework is presented followed by application to limiting cases of various wetting states. The wetting states include those considered in the Wenzel, Cassie-Baxter and wicking state models. The wetting behaviour of multiscale surfaces is then explored by conducting direct simulations of a fluid droplet on a structurally rough surface and a chemically heterogeneous surface. 
\absdiv{\it Findings}
\newline
The underlying origin of the classical wetting models is shown to be rooted within the proposed theoretical framework. In addition, integral geometry provides a topological-based wetting metric that is not contingent on any type of wetting state. Being of geometrical origin the wetting metric is applicable to describe any type of wetting phenomena on complex surfaces. The proposed framework is applicable to any complex fluid topology and multiscale surface.
\end{abstract}


\begin{keywords}
Wetting behavior \sep Geometric state of fluids \sep Topological principles \sep Gaussian curvature \sep Wenzel model \sep Cassie-Baxter model \sep Interfacial curvature 
\end{keywords}

\maketitle{}
\section{Introduction}
Wetting phenomena are familiar in nature and commonly observed in our daily life, and of great significance for basic scientific research as well as providing solutions to cutting-edge technical applications \cite{bonn2009wetting,de1985wetting,de2013capillarity,eller2016operando}. The surface structure and interfacial interactions of liquid and solid are of key importance in determining wetting behavior, and much of the early works have been devoted to studying the forces that influence wetting and spreading behaviors. These forces are studied at the length scale where interfaces are well resolved. The most common system considered is a sessile droplet with an axisymmetric interface on an ideal surface at equilibrium. An increasing number of applications, however, are significantly more complexity than that captured by a sessile drop.

Emerging technologies deal with multiscale surfaces that have curved surfaces, roughness, chemical heterogeneity, complex pore morphology, and a variety of fluid typologies that are not captured by simple wetting models. These technologies include; geological carbon storage \cite{krevor2012relative}, underground hydrogen storage \cite{hashemi2021pore}, fuel cells \cite{meyer2014dead,maheshwari2020nucleation}, high thermal conductivity materials \cite{grosu2020hierarchical}, nano-fluidics \cite{amabili2019pore} and negative compressibility materials \cite{anagnostopoulos2020giant}, which not conform to the constraints of a sessile drop on an ideal surface. For instance, geometries in electro-chemical devices, such as fuel cells and electrolysis cells are geometrically complex to accommodate for maximizing storage volume, transport, and surface area for electro-catalytic conversion \cite{lee2013synchrotron,bazylak2008dynamic,bazylak2009liquid}. The consequence is multiscale hierarchical structures, complex surface morphology, non-axisymmetric interfaces, and complex three-phase contact lines where wetting is difficult to quantify with traditional methods. 

\begin{figure}
\centering\includegraphics[width=0.5\textwidth]{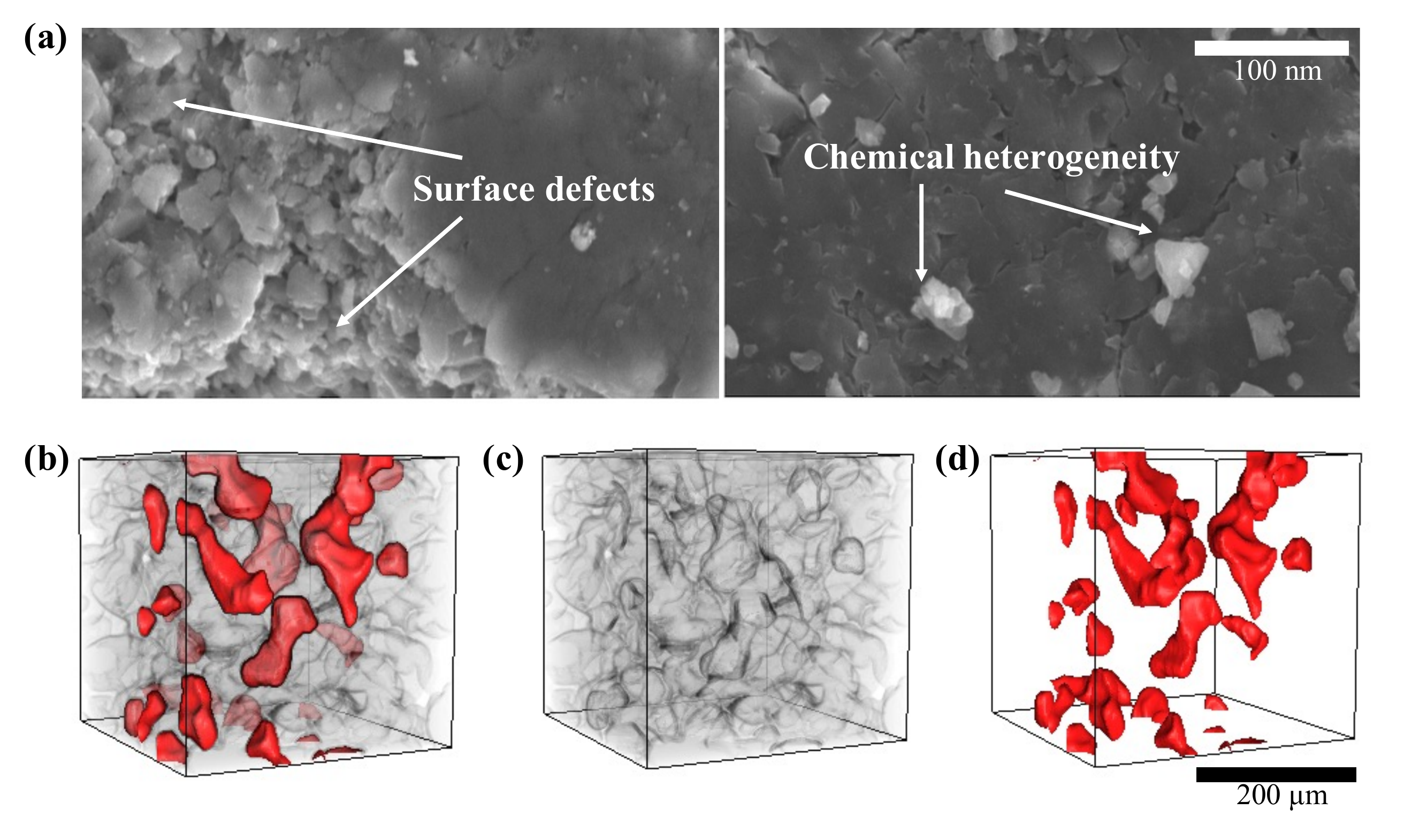}
\caption{(a) The 2D scanning electron microscope (SEM) images with a resolution of $0.5$ $\rm{\mu m}$ for a heterogeneous rock show ubiquitous surface defects and chemical heterogeneity on the solid surface. (b) A three-dimensional (3D) micro-CT image with a dimension of $200 \times 200 \times 200$ voxels for Bentheimer sandstone (grey color) with nonwetting phase (red color) confinement \cite{sun2019bentheimer}. (c) The solid structure of the rock sample. (d) The topology of fluid droplets that are confined in pore space of the rock.}
\label{Topology}
\end{figure}

In the aforementioned technologies, fluid/solid interfaces are often obfuscated by an opaque media that inhibits direct measurement of contact angles. Therefore, traditional wetting studies often focus on the deposition of a fluid droplet on a proxy surface that is similar to that of the porous domain \cite{treiber1972laboratory}. The difficulty in obtaining a proxy surface with structure and chemical composition similar to the actual porous domain is a challenge. A less obvious challenge is the topological aspects of wetting. Within a porous domain, fluids form topologically complex structures \cite{armstrong2016beyond} on rough surfaces, which differ significantly from the sessile droplet structure used for most wettability studies. Scanning electron microscope images of sandstone rock, commonly used in studies for the geological storage of carbon dioxide, display ubiquitous surface defects and chemical heterogeneity on the solid surface, see Figure \ref{Topology}(a). X-ray microcomputed tomography of the same sandstone rock, provided in Figure \ref{Topology}(b,c,d), demonstrate how the topology of the non-wetting phase is confined in the morphologically complex pore space of the rock. Current wetting models do not consider both the multiscale structure of the surface, i.e. sub-scale surface effects, and the topological confinement of the pore structure within a single framework.

Recent works have shown that pore morphology and substrate structure play a significant role in the wetting state \cite{maheshwari2020nucleation}. Spontaneous liquid extrusion form nanopores depends on the pore morphology; a free energy favorable condition occurs for specific conditions of roughness and pore interconnections \cite{amabili2019pore}. For high thermal conductivity materials, a hierarchical pore structure provides superior wetting conditions for antileakage \cite{grosu2020hierarchical}. In contrast to wetting on rigid solids, a wetting ridge, which is a microscopic protrusion caused by the vertical force of the liquid is an important consideration for various applications \cite{soltman2010methodology,gerber2019wetting,furstner2005wetting}. To date, the basic wetting characterization of soft deformable solids is not well understood and several underlying fundamental challenges remain pending \cite{style2012static,snoeijer2018paradox}. The interaction between the liquid interface and the wetting ridge of the solid governs the apparent contact angle. The possible contact angle hysteresis can lead to complex flow phenomena with unusual spreading laws, unstationary behaviors of contact lines or even instabilities with spatial pattern formation that are difficult to characterise with classic wetting models \cite{semprebon2017apparent}.

In previous works, we established a fundamental approach based on integral geometry where wetting behaviour is associated with the Gaussian curvature of a fluid body \cite{sun2020characterization,sun2020probing,armstrong2021multiscale}. Herein, we extend that concept to real-world problems by (1) considering surfaces with multiscale features and (2) explicitly defining fluid topology within the wetting description. We hypothesis that a universal description of wetting on multiscale surfaces can be developed by using integral geometry coupled to thermodynamic laws. We test the theory by considering limiting cases whereby the underlying origin of classic wetting models is shown to be rooted within the presented thermodynamic and integral geometry framework. The wetting behaviour of multiscale surfaces are then explored by conducting direct simulations of a fluid droplet on a structurally rough surface and a chemically heterogeneous surface where contact angle is not constant along the contact line. The proposed framework could be applied to any application were complex fluid topology and multiscale surface effects are expected.

\section{Classic wetting models}
Young's equation explains contact angle for a ideally smooth and homogeneous surface at equilibrium. Most surfaces, however, are not ideal. Wetting on multiscale surfaces presents a more complex and difficult challenge than that for flat and homogeneous surfaces. Non-ideal surfaces are traditionally understood based on the work of Wenzel or Cassie-Baxter \cite{wenzel1936resistance,cassie1944wettability}. These models consider the multiscale nature of a surface where the length scale of the apparent wetting angle can be separated from that of the surface structure. The sub-length scale surface structure is then represented by a single parameter that is representative of the surface. 

\begin{figure}
\centering\includegraphics[width=0.4\textwidth]{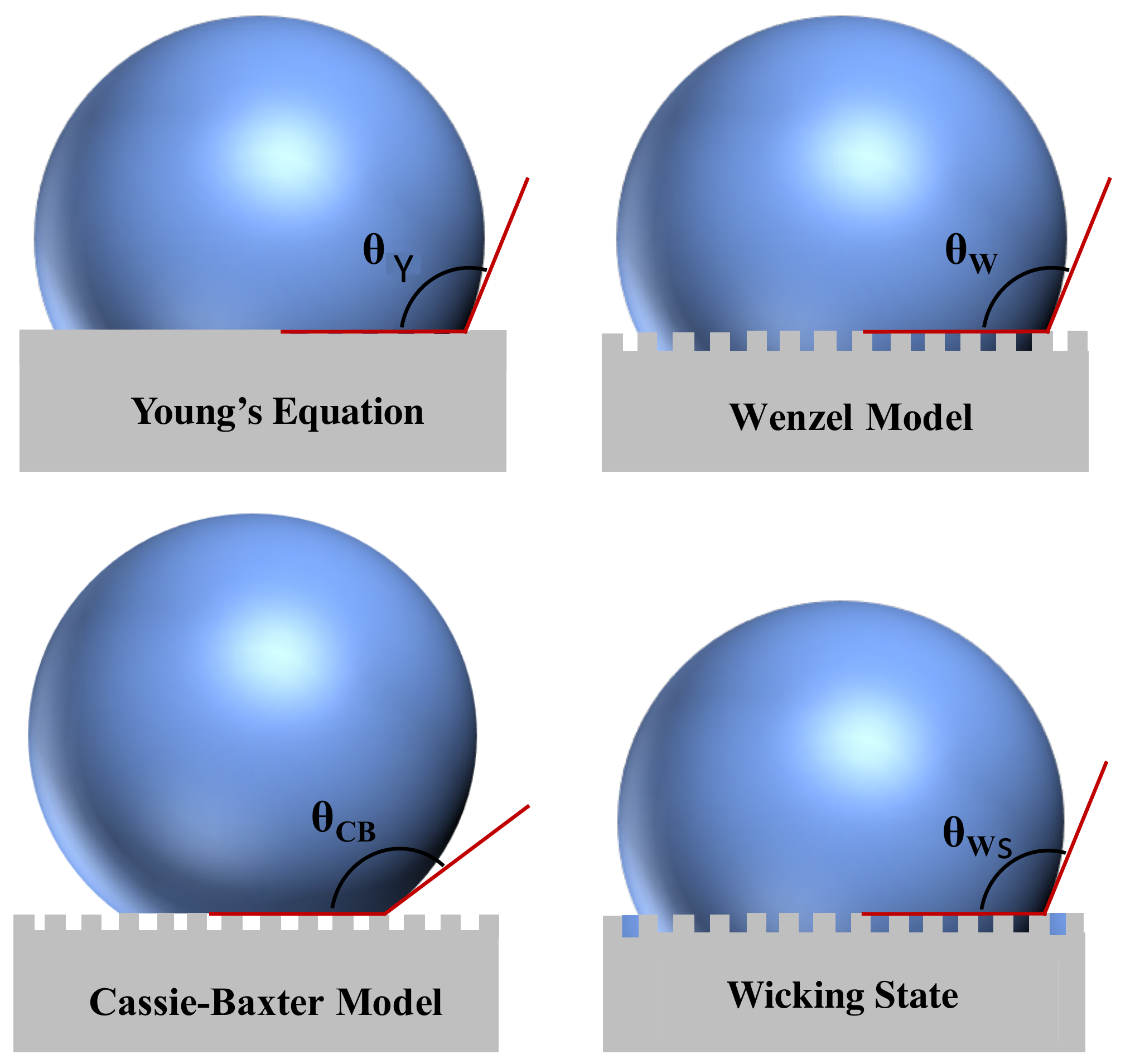}
\caption{Schematic diagrams of liquid droplets deposited on various surfaces to describe the classic wetting models.}
\label{Wenzel_Cassie_Baxter}
\end{figure}

The classic wetting models for non-ideal surfaces are the Wenzel model, Cassie-Baxter model and wicking state model. These models explain an apparent contact angle in terms of the wetting state of the surface and an effective parameter that is representative of the surface. For the Cassie-Baxter model, the fluid droplet sits on top of the surface roughness while the fluid droplet penetrates the roughness in the Wenzel model. In the wicking state model the fluid imbibes or wicks along the roughness ahead of the fluid droplet. Examples of how the fluid wets the surface for each model are provided in Figure \ref{Wenzel_Cassie_Baxter}.

\subsection{Young's equation}
The origin for wetting phenomena is commonly understood based on thermodynamic concepts \cite{wenzel1936resistance,cassie1944wettability,de1985wetting,heslot1990experiments,bonn2009wetting}. Since 1805, Young's equation \cite{doi:10.1098/rstl.1805.0005}, based on thermodynamic laws, is firmly established to infer the wettability of an ideally-flat and chemically homogeneous solid surface at equilibrium, as shown in Figure \ref{Wenzel_Cassie_Baxter}. By considering the surface energies ($\sigma_{ls}, \sigma_{vs}, \sigma_{lv}$) at a microscopic three-phase contact point on an ideally smooth surface, a local force balance leads to Young's equation \cite{young1805iii}

\begin{equation}
\frac{\sigma_{ls} - \sigma_{vs}}{\sigma_{lv}} = \cos \theta_Y,
\label{Young}
\end{equation}
where the subscripts denote the immiscible fluids ($l$, $v$) and solid ($s$). $\theta_Y$ is the contact angle formed along the contact line. While contact angle is interpreted as a geometric measure, it is intrinsically explained in terms of the associated surface energies \cite{bonn2009wetting}. 

\subsection{Wenzel model}
For the Wenzel model, the fluid droplet penetrates the surface roughness as shown in Figure \ref{Wenzel_Cassie_Baxter}. The Wenzel contact angle, $\theta_W$, can be described by

\begin{equation}
\cos \theta_W = r \cos \theta_{Y},
\label{Wenzel}
\end{equation}
where $r$ is the surface roughness factor that is defined as the ratio of the actual area to the projected area of the surface. $\theta_{Y}$ is the intrinsic contact angle governed by Young's equation. It can be deduced from Eq. (\ref{Wenzel}) that surface roughness amplifies the wettability of the original surface. 

\subsection{Cassie-Baxter model}
For the Cassie-Baxter model, the fluid droplet does not penetrate the surface roughness and instead sits on top of the surface. The Cassie-Baxter contact angle, $\theta_{CB}$, can be written as

\begin{equation}
\cos \theta_{CB} = \phi_s \cos \theta_Y + (1-\phi_s),
\label{CB}
\end{equation}
where $\phi_s$ is the area fraction of the liquid/solid interface under the droplet. The surface area fraction under the droplet is important to determine $\theta_{CB}$ since the larger the area fraction of the solid, the larger the apparent contact angle.  

\subsection{Wicking state model}
For super-hydrophilic surfaces, a third type of wetting can occur where the liquid imbibes the surface roughness ahead of the fluid droplet \cite{bico2002wetting}. This type of wetting is called the wicking state. The rough surface can be considered as a type of porous material in which the liquid phase is absorbed. The wicking state contact angle, $\theta_{WS}$, can be written as

\begin{equation}
\cos \theta_{WS} = \phi_d \cos \theta_Y - (1-\phi_d),
\label{WS}
\end{equation}
where $\phi_d$ is the fraction of the solid surface that is dry. For many idealized substrate models $\phi_d$ and $\phi_s$ from the Cassie-Baxter model are equivalent.

\section{Internal energy and fluid topology}
\label{S3}
Our general framework considers the equilibrium condition of a fluid body of arbitrary topology in contact with a multiscale surface of arbitrary geometry. A variational method is used to determine the fluid configuration that minimizes the potential energy. The system energy is explained by Gibbs equation, which is morphologically constrained by a relationship between the total curvature of an object to its topology. The method of Lagrange multipliers is then used to the solve the constrained optimization problem.  

Consider a fluid droplet of arbitrary topology that has a closed interface in a three-phase system, where liquid ($l$), vapor ($v$) and solid ($s$) are present, the total Gaussian curvature of the droplet surface and its global topology are related by the Gauss-Bonnet theorem \cite{chern1944simple}. As a consequence, the Euler characteristic ($\chi$) and its total Gaussian curvature for the manifold of the droplet surface ($M$) obey the following expression,

\begin{equation}
 2\pi \chi(M)=\int_{M} \kappa_G dA + k_d,
 \label{GBT}
\end{equation}
where $dA$ is the Riemannian area element of droplet interface along the surface manifold $M$, $\kappa_G = 1/(r_1r_2)$ is the Gaussian curvature along the surface area element $dA$, $r_1$ and $r_2$ are the two principal radii of curvature at any point on the surface, and $k_d$ is deficit curvature \cite{sun2020linking}. 

\begin{figure}
\hspace*{-0.1cm}\includegraphics[width=0.5\textwidth]{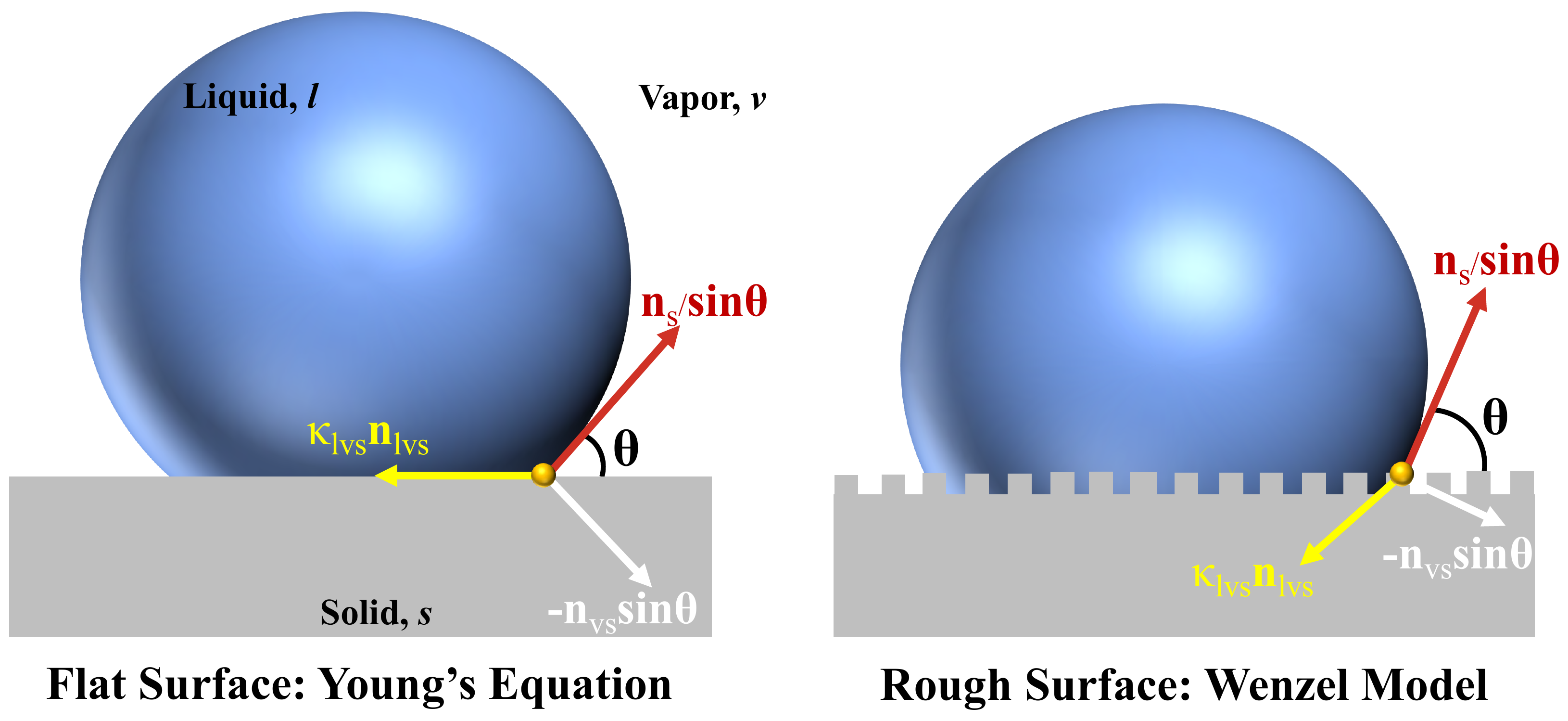}
\caption{The 2D schematics of a fluid droplet immersed within an immiscible vapor phase ($v$) depositing on solid ($s$) substrate with and without surface roughness. The labelled definitions definitions corresponding to Eq. (\ref{eq:kd-explicit}).}
\label{2D_label}
\end{figure}

The deficit curvature, $k_d$, can be further defined as $\int_{\partial M} \kappa_g dC$, where $dC$ is the Riemannian line element along the contact line $\partial M$ formed by the contact of the droplet with the solid surface and $\kappa_g$ is the geodesic curvature along the contact line $\partial M$, which is the curvature of the projection of $\partial M$ onto the tangent plane to the surface manifold. These regions correspond physically to the contact line. Based on this, the deficit curvature can be written explicitly as an integral measure of the local contact angle,

\begin{equation}
k_d = \int_{\partial M} \kappa_{lvs} \bm{n}_{lvs} \cdot 
\big[ \bm{n}_{s}  \sin \theta 
+ \bm{n}_{vs}(1 - \cos \theta)  \big] dC,
\label{eq:kd-explicit}
\end{equation}
where 
$\kappa_{lvs}$ is the curvature of the contact line,
$\bm{n}_{lvs}$ is the normal vector in the direction of the contact
line curvature, 
$\bm{n}_{s}$ is the normal vector relative to the solid surface,
and 
$\bm{n}_{vs}$ is the outward normal vector to the contact line
within the tangent plane for the solid surface. Figure \ref{2D_label} provides reference to the aforementioned parameters for both flat and rough surfaces. With a flat surface, all curvature for the contact line lies within the solid plane, so $\kappa_{lvs}\bm{n}_{lvs}$ also lies within the plane. For a rough (or curved) surface, the solid roughness (or surface curvature) allows the contact line to curvature in a variety of directions, and thus $\kappa_{lvs}\bm{n}_{lvs}$ does not lie within the solid plane. For further details on the reference frame used see the Darboux frame in Euclidean space \cite{guggenheimer1977differential}. 

Eq. (\ref{eq:kd-explicit}) results from the projection of the geodesic curvature for the non-smooth parts of
$M$ onto the reference frame for the solid surface, 
which is used to compute the contact angle.
Deficit curvature therefore provides a way to directly upscale the 
contact angle, which is particularly useful
in situations where contact angle varies due to heterogeneity. This provides a general description of a fluid droplet of an arbitrary topology.

When upscaling the contact angle, a macroscopic contact angle,$\theta^{macro}$, can be defined based on deficit curvature.

\begin{equation}
\theta^{macro}=\lambda \frac{k_d}{N_c},
\label{normal_single}	
\end{equation}
where $N_c$ is the number of contact line loops and $\lambda$ is a scale factor. Normalization is performed to account for the fact that the deficit curvature is an extensive
property of a system, since the Euler characteristic is an additive quantity. Therefore, it is intuitive to define an effective property so that the macroscopic contact angle is defined on the interval $|\theta^{macro}| \in [0,\pi]$, which occurs with $\lambda =4$, as shown in previous work \cite{sun2020probing}. For a sessile drop system, $N_c = 1$, while other more complex systems can be considered as well \cite{sun2020characterization}. 

At equilibrium, the shape of a fluid droplet is determined based on minimization of the surface free energy, which in turn determines the shape of the contact line. We will assume that the internal energy, $U$, of a fluid droplet system is well-approximated by

\begin{equation}
\begin{split}
U = TS - p_l V_l - p_v V_v + \sigma_{lv} A_{lv}
+ \sigma_{vs} A_{vs} + \sigma_{ls} A_{ls} \;,
\end{split}
\label{Gibbs}
\end{equation}
where $\sigma$ is surface energy, $p$ is pressure and $V$ is volume. 
The variation of the internal energy is given by

\begin{equation}
\begin{split}
\delta U = T \delta S - p_l \delta V_l - p_v\delta V_v + \sigma_{lv} \delta A_{lv} \\
+ \sigma_{vs}\delta A_{vs} + \sigma_{ls} \delta A_{ls} \;.
\end{split}
\label{Gibbs_var}
\end{equation}
Based on this, we consider a closed system with $\delta U = 0$ and constant volume. Fluid rearrangements 
can occur provided that the volume of the droplet is not changed, meaning that
$\delta V_a = \delta V_b = 0$. Subject to these restrictions, entropy change in the system is strictly linked to minimization of the surface energy.

In the following, we will consider limiting cases of Eqs. (\ref{GBT}) and (\ref{Gibbs}) to provide the well-known Young's equation, Wenzel model, Cassie-Baxter model and wicking state model. Other more general cases can also be considered depending on the particular wetting system of interest. We will start by defining the geometrical expressions for a sessile drop system as used in the classic models.  

\subsection{Geometrical expressions}
\begin{figure}
\centering\includegraphics[width=0.4\textwidth]{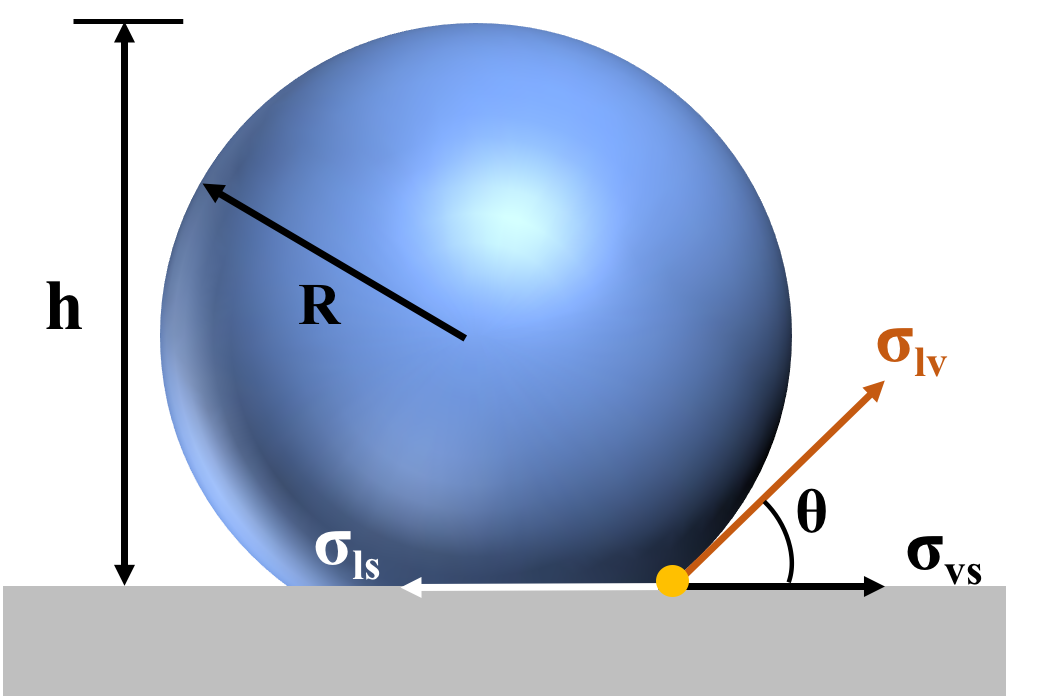}
\caption{The labelled notations of a droplet deposited on an solid substrate to illustrate the derivation of the Wenzel, Cassie-Baxter and wicking state models.}
\label{Young_derivation}
\end{figure}

We consider a length scale at which a fluid droplet on a solid substrate is well resolved while features on the solid substrate are not resolved. The surface can have sub-length scale roughness or chemical heterogeneity, and these features are represented by a single value at the length scale of the fluid droplet. This is the commonly used "representative elementary volume" assumption applied in continuum mechanics \cite{bear2013dynamics}. The system is displayed in Figure \ref{Young_derivation}. 

The Gauss-Bonnet theorem for the fluid droplet, $D$, as displayed in Figure \ref{Young_derivation} is

\begin{equation}
4 \pi \chi(D) = k_d + \kappa_{lv} A_{lv} + \kappa_{ls} A_{ls}, \;
\label{GB_droplet}
\end{equation}
where $\chi$ is the Euler characteristic, $\kappa$ is curvature and subscripts $d$, $lv$, and $ls$ denote deficit curvature, the liquid/vapour interface and the liquid/solid interface, respectively.
In addition, even though the substrate surface could be rough the curvature of the surface at the length scale of the fluid droplet is flat for the example in Figure \ref{Young_derivation}, and thus $\kappa_{ls} = 0$ would apply. Euler characteristic is a scale invariant property, and thus would not change if the droplet was observed at a length scale at which the surface roughness was resolved. Therefore, the curvature due to roughness of the solid/liquid interface is not considered in Eq. (\ref{GB_droplet}) but could be incorporated for surfaces that have significant curvature at the length scale of the fluid droplet. In the following, we will consider the corresponding thermodynamic equilibrium based on the assumption of constant topology, $\chi(C) = \mbox{constant}$. Other variations of constant fluid topology could also be considered, such as a torus or any other $\chi =  \mbox{constant}$ morphology, as observed in porous media systems.

Geometrical expressions can be computed by considering a spherical cap with droplet radius $R$ and droplet height $h$ to the solid surface:

\begin{equation}
\begin{split}
k_d &= 2 \pi (1-\cos \theta),  \\
\kappa_{lv} &= \frac{1}{R^2}, \\
A_{lv} &= 2 \pi R h, \\
A_{ls} &= \pi h (2R-h).
\end{split}
\label{eq:12} 
\end{equation}
The associated variations are:

\begin{equation} \label{eq:13}
\begin{split}
\delta k_d &= \delta \big [ 2 \pi (1-\cos \theta) \big ] 
             = 2 \pi (h R^{-2}\delta R - R^{-1} \delta h ),  \\
        \delta \kappa_{lv} &= \delta \big [ R^{-2} \big] = -2 R^{-3} \delta R, \\
\delta A_{ls} &= \delta \big [\pi h (2R-h) \big] =
                    \pi \big( 2 h \delta R + 2 R \delta h - 2 h \delta h \big). 
\end{split}
\end{equation}
The volume of a spherical cap is $V = \frac 1 3 \pi h^2 \big( 3 R-h \big)$ and setting $\delta V=0$ imposes the relationship $\delta R = \Big( 1 - \frac{2R}{h} \Big) \delta h$. Inserting this into the Eqs. (\ref{eq:13}) provides:

\begin{equation}
\begin{split}
\delta k_d &= 2 \pi \Big(\frac{h}{R^2} - \frac{3}{R} \Big) \delta h, 
\\
\delta \kappa_{lv} &= \Big(\frac{4}{R^2h} - \frac{2}{R^3} \Big) \delta h, \\
\delta A_{ls} &=  -2 \pi R \delta h.
\end{split}
\label{eq:14}
\end{equation}

\subsection{Limiting case: Young's equation}
The derivation of Young's equation was presented in \cite{sun2020characterization} and repeated here for completeness. Young's equation considers a single fluid droplet sitting on a smooth and homogeneous surface. The variation of the internal energy at constant fluid volume is

\begin{multline}
\delta U = T \delta S + \sigma_{lv} \delta A_{lv} + \sigma_{ls}\delta A_{ls} + \sigma_{vs} \delta A_{vs} \;.
\label{eq:3}
\end{multline}
We consider a closed system with $\delta U = 0$. 
The maximum entropy corresponds to the minimum for the surface
energy,

\begin{equation}
\delta S = - \frac 1 T \Big[ \sigma_{lv} \delta A_{lv}
+ \sigma_{ls}\delta A_{ls} + \sigma_{vs} \delta A_{vs} \Big] \ge 0 \;.
\label{eq:4}
\end{equation}
If the total solid surface area is constant, then 

\begin{equation}
\delta A_s = \delta A_{vs} + \delta A_{ls} = 0 \;,
\label{eq:5}
\end{equation}
which can be used to eliminate one of the surface areas from Eq. (\ref{eq:4}). 

We can also use the topological constraint from Eq. (\ref{GB_droplet}) to determine 
the condition that should be imposed to ensure that geometric variation 
occurs without changing the topology, which provides

\begin{multline}
\delta \big[ 4 \pi \chi(D)\big] = \delta k_d + 
\kappa_{lv} \delta A_{lv} + A_{lv} \delta \kappa_{lv}  = 0
 \;.
\label{eq:6}
\end{multline}
This expression is a simple statement that while the total curvature of the object
cannot change, it can be redistributed along the boundary. We shall impose Eq. 
(\ref{eq:6}) as a constraint on Eq. (\ref{eq:4}) using the method of Lagrange multipliers,
also using Eq. (\ref{eq:5}) to eliminate $\delta A_{vs}$

\begin{multline}
\delta S = - \frac 1 T \Big[ \sigma_{lv} \delta A_{lv}
+ \big( \sigma_{ls} - \sigma_{vs} \big) \delta A_{ls} \Big] \\
+ \lambda \Big[ \delta k_d + 
\kappa_{lv} \delta A_{lv} + A_{lv} \delta \kappa_{lv} \Big].
\end{multline}
To eliminate $\delta A_{lv}$, we choose 

\[
\lambda = \frac{\sigma_{lv}}{\kappa_{lv} T} \;.
\]
The entropy change can now be expressed as

\begin{multline}
    \delta S = - \frac{1}{T} \Big[ (\sigma_{ls}-\sigma_{vs}) \delta A_{ls} \\ +\frac{\sigma_{lv}}{\kappa_{lv}}(\delta k_d + A_{lv}\delta \kappa_{lv}) \Big].
 \label{eq:7}
\end{multline}
The first term corresponds to variations in surface area while the second term corresponds to the redistribution of the total curvature that occurs at constant surface area. One of the ways to redistribute the total curvature
is by altering the deficit curvature, which changes the contact angle. This provides the link between deficit curvature and wetting \cite{sun2020probing}. Since $T>0$ the inequality can be re-expressed in the form

\begin{multline}
\Big( \frac{\sigma_{ls} - \sigma_{vs}}{\sigma_{lv}} \Big) \kappa_{lv} \delta A_{ls}
+ \delta k_d + A_{lv} \delta \kappa_{lv}  \le 0
 \;.
 \label{eq:8}
\end{multline}
This expression demonstrates that the surface energy of the system must decrease. Inserting $\kappa_{lv}$ and $A_{lv}$ from Eqs. \ref{eq:12}, $\delta k_d$ and $\delta A_{ls}$ from Eq. (\ref{eq:14}), followed by simplification provides 

\begin{eqnarray}
 \frac{2\pi \delta h }{R} \Bigg\{
1 - \frac{h}{R}  - \frac{\sigma_{ls} - \sigma_{vs}}{\sigma_{lv}} 
\Bigg\}\le 0\;.
\end{eqnarray}
Noting that $cos \theta = 1 - h/R$, it can be observed that for a general variation $\delta h$
and $R>0$, maximum entropy will be obtained based on the condition that

\begin{align}
    \cos \theta - \frac{\sigma_{ls} - \sigma_{vs}}{\sigma_{lv}}  = 0.
\end{align}
This is Young's equation.

\subsection{Limiting case: Wenzel's model}
The Wenzel model considers a rough surface where the liquid phase penetrates the surface roughness, and thus increases the contact area between the liquid and solid phases. The additional liquid/solid contact area due to roughness is accounted for by $r$. Therefore the geometrical expression for the liquid/solid surface area is

\begin{equation}
A_{ls}= \pi r h (2R-h),
\end{equation}
and the associated variation is 

\begin{equation}
\delta A_{ls} = -2 \pi r R \delta h.
\label{roughness_var}
\end{equation}
Considering the same conditions and steps as for Young's equation, we again arrive at the following inequality

\begin{multline}
\Big( \frac{\sigma_{ls} - \sigma_{vs}}{\sigma_{lv}} \Big) \kappa_{lv} \delta A_{ls}
+ \delta k_d + A_{lv} \delta \kappa_{lv}  \le 0
 \;.
 \label{inequality_Wenzel}
\end{multline}
Inserting $\kappa_{lv}$ and $A_{lv}$ from Eqs. (\ref{eq:12}), $\delta k_d$ from Eq. (\ref{eq:14}) and $\delta A_{ls}$ from Eq. (\ref{roughness_var}), followed by simplification provides 

\begin{eqnarray}
 \frac{2\pi \delta h }{R} \Bigg\{
1 - \frac{h}{R}  - r \frac{\sigma_{ls} - \sigma_{vs}}{\sigma_{lv}} 
\Bigg\}\le 0\;.
\end{eqnarray}
Noting that $cos \theta = 1 - h/R$, it can be observed that for a general variation $\delta h$
and $R>0$, maximum entropy will be obtained based on the condition that

\begin{align}
    \cos \theta -r\frac{\sigma_{ls} - \sigma_{vs}}{\sigma_{lv}}  = 0.
\end{align}
Lastly, applying Young's equation, we arrive at

\begin{equation}
\cos \theta = r \cos \theta_Y, 
\end{equation}
which is Wenzel's model.

\subsection{Limiting case: Cassie-Baxter model}
The Cassie-Baxter Model considers a rough surface where the liquid droplet sits on top of the roughness. For this model the solid fraction in contact with the liquid must be considered. The variation of the internal energy for this wetting state at constant fluid volume is

\begin{multline}
\delta U = T \delta S + \sigma_{lv} \delta A_{lv} + \sigma_{lv}(1-\phi_s)\delta A_{ls} \\
+ \sigma_{ls} \phi_s \delta A_{ls} + \sigma_{vs}\phi_s\delta A_{vs}\;.
\label{CB_energy_var}
\end{multline}
Considering a closed system and with further simplification provides 

\begin{multline}
\delta S = - \frac 1 T \Big [\sigma_{lv} \delta A_{lv} +\big( (\sigma_{ls}-\sigma_{vs})\phi_s \\
+ \sigma_{lv} (1-\phi_s)
\big) \delta A_{ls} \Big ] \ge 0 \;.
\label{CB_S}
\end{multline}
Eq. (\ref{eq:6}) can be imposed as a constraint on Eq. (\ref{CB_S}) using the method of Lagrange multipliers to express the entropy change as 

\begin{multline}
    \delta S = - \frac{1}{T} \Big[ (\sigma_{ls}-\sigma_{vs}) \phi_d \delta A_{ls} + \sigma_{lv}(1- \phi_s)
    \delta A_{ls} + \\
    \frac{\sigma_{lv}}{\kappa_{lv}}(\delta k_d + A_{lv} \delta \kappa_{lv}) \Big]
 \label{Entropy_GB}.
\end{multline}
Furthermore, since $T>0$ the inequality can be re-expressed in the form

\begin{multline}
   \Big [ \frac{(\sigma_{ls}-\sigma_{vs})}{\sigma_{lv}} \phi_s + (1-\phi_s)
    \Big] \kappa_{lv} \delta A_{ls} \\
    + \delta k_d + A_{lv} \delta \kappa_{lv} \le 0
 \label{CBM}.
\end{multline}
Applying Eqs. (\ref{eq:12}) and (\ref{eq:14}) to Eq. (\ref{CBM}) followed by simplification provides

 \begin{eqnarray}
 \frac{2\pi \delta h }{R} \Bigg\{
\cos \theta  -\phi_s \frac{\sigma_{ls} - \sigma_{vs}}{\sigma_{lv}} 
 - (1-\phi_s) \Bigg\}\le 0\;.
\end{eqnarray}
By applying Young's equation and observing that for a general variation $\delta h$
and $R>0$, maximum entropy is obtained based on the condition that

\begin{equation}
\cos \theta = \phi_s \cos \theta_Y + (1-\phi_s).
\end{equation}
This is the Cassie-Baxter model.

\subsection{Limiting case: wicking state model}
The wicking state model considers the fraction of the solid surface that is dry, which is the fraction of the solid surface that contacts the vapour phase. The variation of the internal energy for the wicking state at constant fluid volume is

\begin{multline}
\delta U = T \delta S + \sigma_{lv} \delta A_{lv}
+ \sigma_{ls} \phi_d \delta A_{ls} \\
+ \sigma_{vs} \phi_d\delta A_{vs} + \sigma_{lv}(1-\phi_d)\delta A_{vs}\;.
\label{Gibbs_wicking}
\end{multline}
Considering a closed system and with further simplification provides

\begin{multline}
\delta S = - \frac 1 T \Big[ \sigma_{lv} \delta A_{lv}+ (\sigma_{ls}-\sigma_{vs})\phi_d \delta A_{ls} \\
- \sigma_{lv} (1-\phi_d) \delta A_{ls}
\Big] \ge 0 \;.
\label{wicking_s}
\end{multline}
Eq. (\ref{eq:6}) can be imposed as a constraint on Eq. (\ref{wicking_s}) using the method of Lagrange multipliers to express the entropy change as 

\begin{multline}
    \delta S = - \frac{1}{T} \Big[ (\sigma_{ls}-\sigma_{vs}) \phi_d \delta A_{ls} - \sigma_{lv}(1-\phi_d)
    \delta A_{ls} \Big] \\
    +  \frac{\sigma_{lv}}{\kappa_{lv}}(\delta k_d + A_{lv} \delta \kappa_{lv}) \Big]
 \label{Entropy_WS}.
\end{multline}
Furthermore, since $T>0$ the inequality can be re-expressed in the form

\begin{multline}
   \Big [ \frac{(\sigma_{ls}-\sigma_{vs})}{\sigma_{lv}} \phi_d - (1-\phi_d)
    \Big] \kappa_{lv} \delta A_{ls} \\
    + \delta k_d + A_{lv} \delta \kappa_{lv} \le 0
 \label{wicking}.
\end{multline}
The same geometrical expressions and associated variations as used in the previous section for the Cassie-Baxter model also apply here. Applying the geometrical expressions to Eq. (\ref{wicking}) followed by simplification provides 

 \begin{eqnarray}
 \frac{2\pi \delta h }{R} \Bigg\{
\cos \theta  -\phi_d \frac{\sigma_{ls} - \sigma_{vs}}{\sigma_{lv}} 
 + (1-\phi_d) \Bigg\}\le 0\;.
\end{eqnarray}
By applying Young's equation and observing that for a general variation $\delta h$
and $R>0$, maximum entropy will be obtained based on the condition that

\begin{equation}
\cos \theta = \phi_d \cos \theta_Y - (1-\phi_d).
\end{equation}
This is the wicking state model.

\section{Wetting on multiscale surfaces}
The derived models demonstrate the universal principle of energy minimization within a specific topological constraint. The Gauss-Bonnet theorem describes how curvature is distributed throughout the fluid droplet. The geodesic curvature of the contact line is defined as, $k_d$, being the curvature required within the contact line to complete the object. As demonstrated in the derivations, the contact angle changes with the redistribution of total curvature to $k_d$ providing a link between these two measures. Defining the wetting state of an object by $k_d$ has an advantage over contact angle for cases where the contact angle is not constant along the contact line. On a multiscale surface the contact angle will be sensitive to sub-scale properties, varying based on local surface chemistry and roughness when the length scale of observation is small enough. Alternatively, $k_d$ provides a single value that accounts for the sub-scale variations. While contact angle depends on the length scale of the observation, topological measures are scale invariant.

In the following, simulations are conducted where a fluid droplet is resolved along with the sub-scale substrate features. The local variation of contact angle is assessed along with $k_d$. The simulations are conducted to provide justification for using $k_d$ as a wetting descriptor within the integral geometry framework presented in Section \ref{S3}.

\subsection{Simulations}
Quasi-steady-state simulations of a 3D sessile oil droplet were performed on various types of solid surface to investigate the wetting behavior on multiscale surfaces and to validate the usage of deficit curvature as a topological constraint. The shape of a sessile droplet on a substrate is governed by the Young-Laplace equation, which balances the surface tension and internal pressure. A variational approach was adopted to successively minimizing the overall energy of the drop to obtain a final equilibrium shape, which was achieved by using the Surface Evolver software \cite{brakke1992surface}. 

Surface Evolver is developed based on the principle of minimization of energy and conservation of volume subjected to constraints. The surface is evolved to a minimum energy via a gradient descent method. In a typical multiphase system, its overall energy is the sum of its interfacial potential energy. The interfacial potential energy of the sessile droplet can be restricted to the sum of interfacial energies as

\begin{equation}
    E=\int\int_{A_{ls}}\sigma_{ls}dA+\int\int_{A_{lv}}\sigma_{lv}dA+\int\int_{A_{vs}}\sigma_{vs}dA,
\end{equation}
where $\sigma$ and $A$ represent surface tension and surface area, respectively. By applying Young's equation, the interfacial potential energy becomes

\begin{equation}
    E=\sigma_{lv}[A_{lv}-\int\int_{A_{ls}}\cos \theta_{Y} dA].
\end{equation}

\begin{figure}
\centering\includegraphics[width=0.5\textwidth]{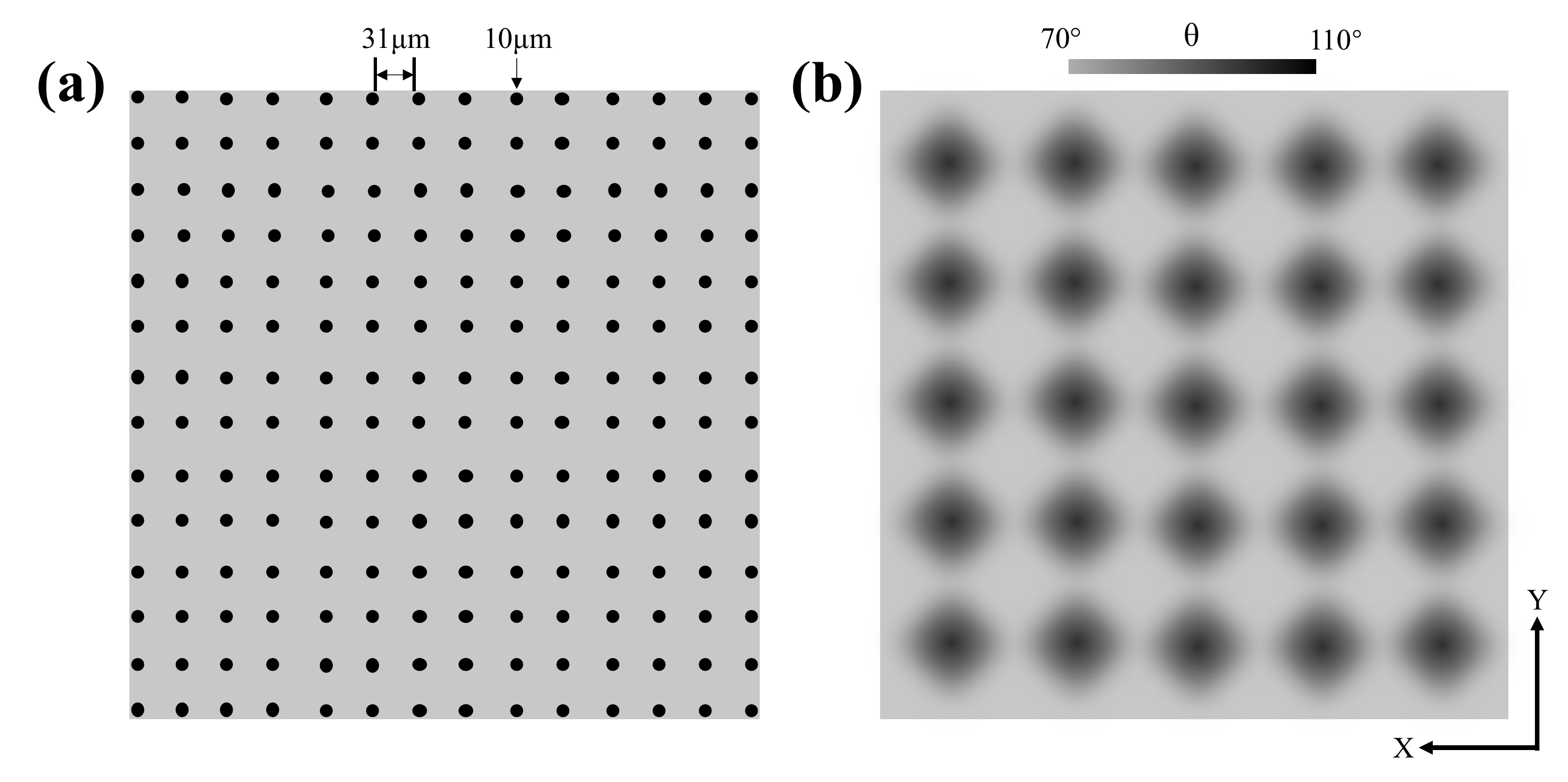}
\caption{(a) The pillar (red color) arrangement of structural patterned surface with a pillar diameter of $10$ $ \mu m$ and a pillar width of $31$ $ \mu m$. (b) The variation of wettability based on Eq. (\ref{chem_vary}) for chemical heterogeneous surface where the black color represents the maximum contact angle assigned to the surface and the grey color represents the minimum contact angle assigned to the surface.}
\label{pattern}
\end{figure}

For patterned surface generation, the vertices of pillar facets are constrained such that their $z$-coordinates are zero corresponding to the pillar surface. The $x$- and $y$-coordinates of those vertices are constrained to lie on the boundary of the pillars. The coordinates of the vertices belonging to the liquid/vapor interface are not constrained. In other words, the facets of the liquid/solid interface are forbidden to deform beyond their perspective pillar boundaries but the remaining droplet facets are unconstrained. The resulting pillar arrangement used is depicted in Figure \ref{pattern}(a) with $10$ $\mu m$ in diameter pillars and $31$ $\mu m$ spacing between the pillars.

For the generation of a chemically heterogeneous surface, the wettability of the surface is designed as a 2D periodic pattern that is controlled by the energy function $f(x,y)$ in $x$- and $y$-coordinates of the solid surface as

\begin{multline}
f(x,y)= \cos(\theta_{Y}) + \\
\psi [cos(4\pi x)cos(4\pi y)-cos(4\pi x)-cos(4\pi y)],
\label{chem_vary}
\end{multline}
where $\theta_{Y}$ is the intrinsic contact angle from Young's equation and has a value of $90^\circ$. $\psi$ is a weight factor to control the maximum advancing contact angle and minimum receding contact angle. A value of $0.2$ is applied in this study. A schematic illustration of wettability variation for the chemical heterogeneous surface is shown in Figure \ref{pattern}(b). As a result, the maximum advancing angle and minimum receding angle are between $110^\circ$ and $70^\circ$, respectively. This tens of degrees variation in apparent contact angles is often attributed to the difference in direct ionic bonding of the topmost molecular layer in chemical heterogeneous surfaces \cite{wang2011probing}, which is known to have an impact on the vicinity of the contact line. Consequently, it leads to the change in droplet interfacial curvature and thereby the resulting wetting hysteresis. 

\subsection{Results}
The simulation results are used to investigate the nature of wetting on multiscale surfaces. Firstly, the wetting behaviour is investigated at the length scale at which the surfaces features are fully resolved by measuring a local microscopic contact angle at each point along the contact line. Secondly, the wetting behaviour is investigated by considering various ways to measure an effective contact angle. We calculate the Cassie-Baxter contact angle,$\theta_{CB}$, as defined in Eq. (\ref{CB}) and apply the concept of deficit curvature to measure a macroscopic contact angle, $\theta^{macro}$. In addition, an apparent contact angle,$\theta_{app}$, is determined by fitting the entire sessile droplet to a sphere, and then computing the effective Young-Laplace equivalent angle, which could be the angle measured by experimental observation. 

\begin{figure}
\centering\includegraphics[width=.5\textwidth]{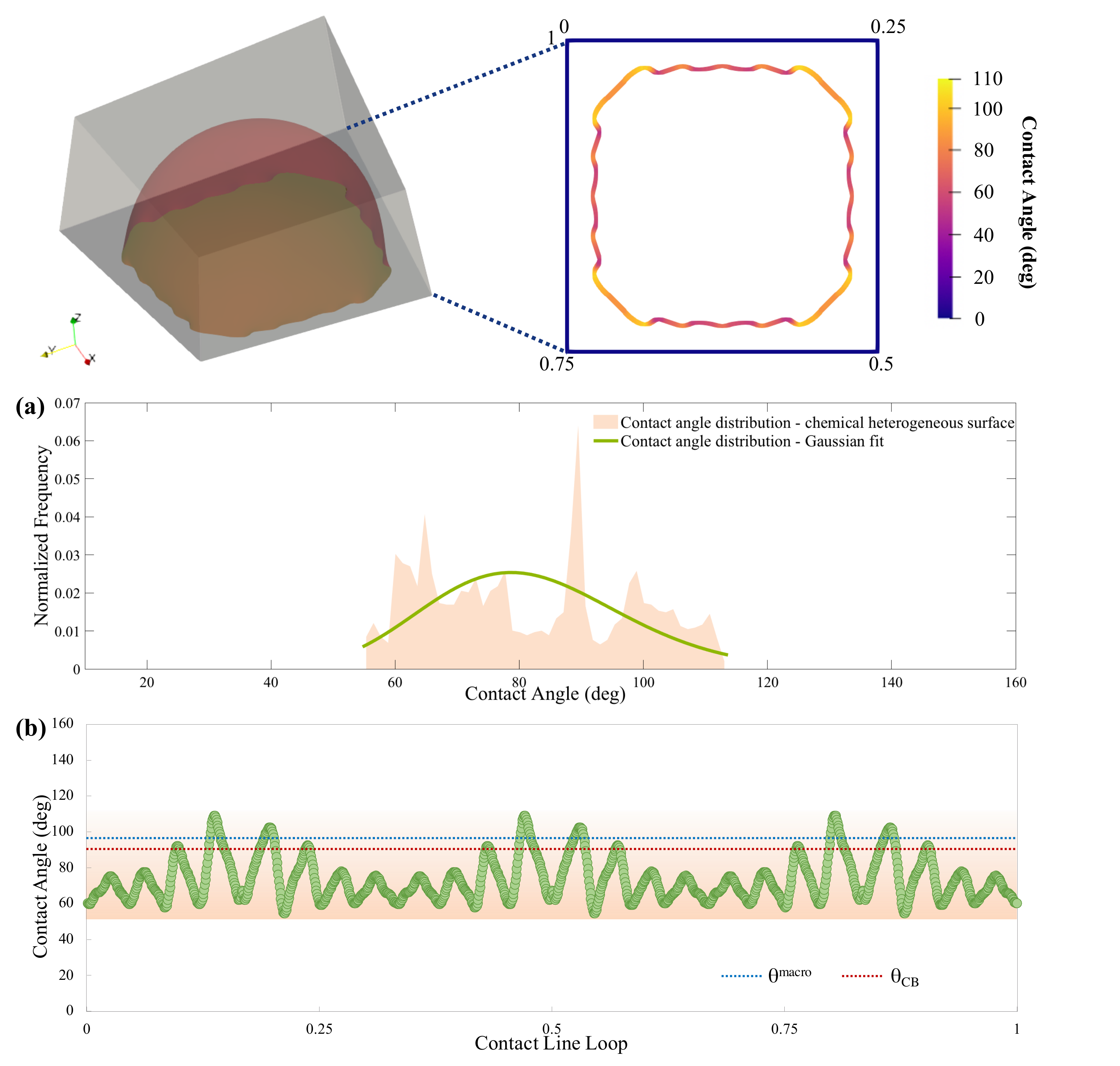}
\caption{Simulation of 3D oil droplet on chemical heterogeneous surface that shows the three-phase contact points with associated local contact angle values. (a) The local contact angle distribution for chemical heterogeneous surface with its Gaussian fit. (b) The hysteresis loop of contact angles from the beginning ($0$) to the end ($1$) of the contact line and the computed $\theta^{macro}$ and $\theta_{CB}$ values. The shaded area is the range between the advancing and receding contact angles.}
\label{chem}
\end{figure}

\begin{figure}
\centering\includegraphics[width=0.5\textwidth]{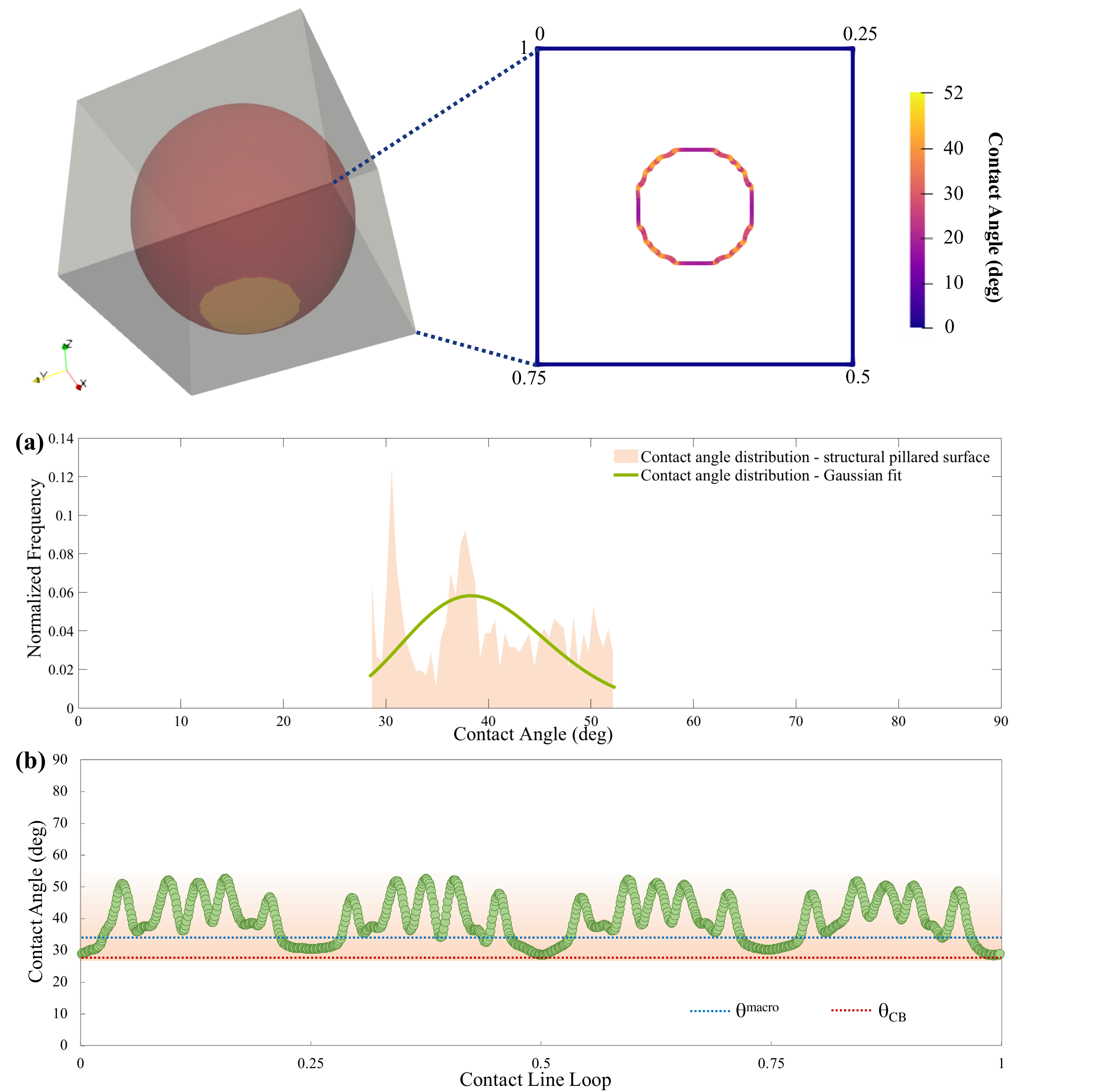}
\caption{Simulation of 3D oil droplet on structural pillared surface that shows the three-phase contact points with associated local contact angle values. (a) The local contact angle distribution for structural pillared surface with its Gaussian fit. (b) The hysteresis loop of contact angles from the beginning ($0$) to the end ($1$) of the contact line and the computed $\theta^{macro}$ and $\theta_{CB}$ values. The shaded area is the range between the advancing and receding contact angles.}
\label{rough}
\end{figure}

In Figure \ref{chem} and Figure \ref{rough}, the variation of contact angle along the contact line is observed. This variation causes difficulty in characterizing wettability in complex and confined porous media for direct numerical simulation of fluids flow as well as the design of functional surfaces since it is unclear which value of contact angle best represents the wetting state of the system. Conversely, both $\theta_{CB}$ and $\theta^{macro}$ provide an apparent angle that represents the local variations. 

As shown in Figure \ref{chem}, the fluid droplet has a volume of $0.2$ $\rm \mu L$ immersed in another immiscible phase and deposited on a chemically heterogeneous surface (Figure \ref{pattern}(b)) with a periodic pattern of wettability near $90^{\circ}$. In Figure \ref{chem}, the computed $\theta^{macro}$ for the droplet is $98.4^{\circ}$. In addition, the hysteresis loop of the local contact angles is presented in Figure \ref{chem}(b) and the mean value is $81.8^{\circ}$, which provides the spatially varying and apparent wetting information at the microscale. Since the wetting state of the surface varies periodically in the $x$- and $y$-coordinates as stated in Eq. (\ref{chem_vary}), the associated surface free energy varies as well. To compute the area fraction ($f_i$) and then $\theta_{CB}$, the contact area between the droplet and solid surface is segmented into small regions ($i$). Furthermore, an average value ($\theta_i$) is assigned to each small area region based on the energy function of Eq. (\ref{chem_vary}). $\theta_{CB}$ can thereby be computed by applying Eq. (\ref{CB}). A value of $90.1^{\circ}$ is thereby obtained for $\theta_{CB}$ when $i = 0.01$. As observed, $\theta^{macro}$ provides a value that is slightly larger than $\theta_{CB}$, which represents the macroscopic wetting behavior of the droplet on the chemical homogeneous surface and provides a representation of the local measures. 

For the droplet deposited on the micro-grooved surface as shown in Figure \ref{rough}, an intrinsic contact angle of $60^{\circ}$ was set for the surface. The droplet is constrained on the solid surface providing the wetting fraction required for Eq. (\ref{CB}). In Figure \ref{rough}, $\theta^{macro}$ is $34.5^{\circ}$, while the mean value of the microscopic contact angle measurements along the contact line is $39.5^{\circ}$. In addition, ($\theta_{CB}$) is $28.1^{\circ}$. Similar to the results for chemical heterogeneous surface, the observed value of $\theta^{macro}$ provides a value that is larger than the $\theta_{CB}$, while the value of $\theta_{CB}$ is noticeably close to the advancing angle for the oil droplet. 

\begin{figure}
\centering\includegraphics[width=0.47\textwidth]{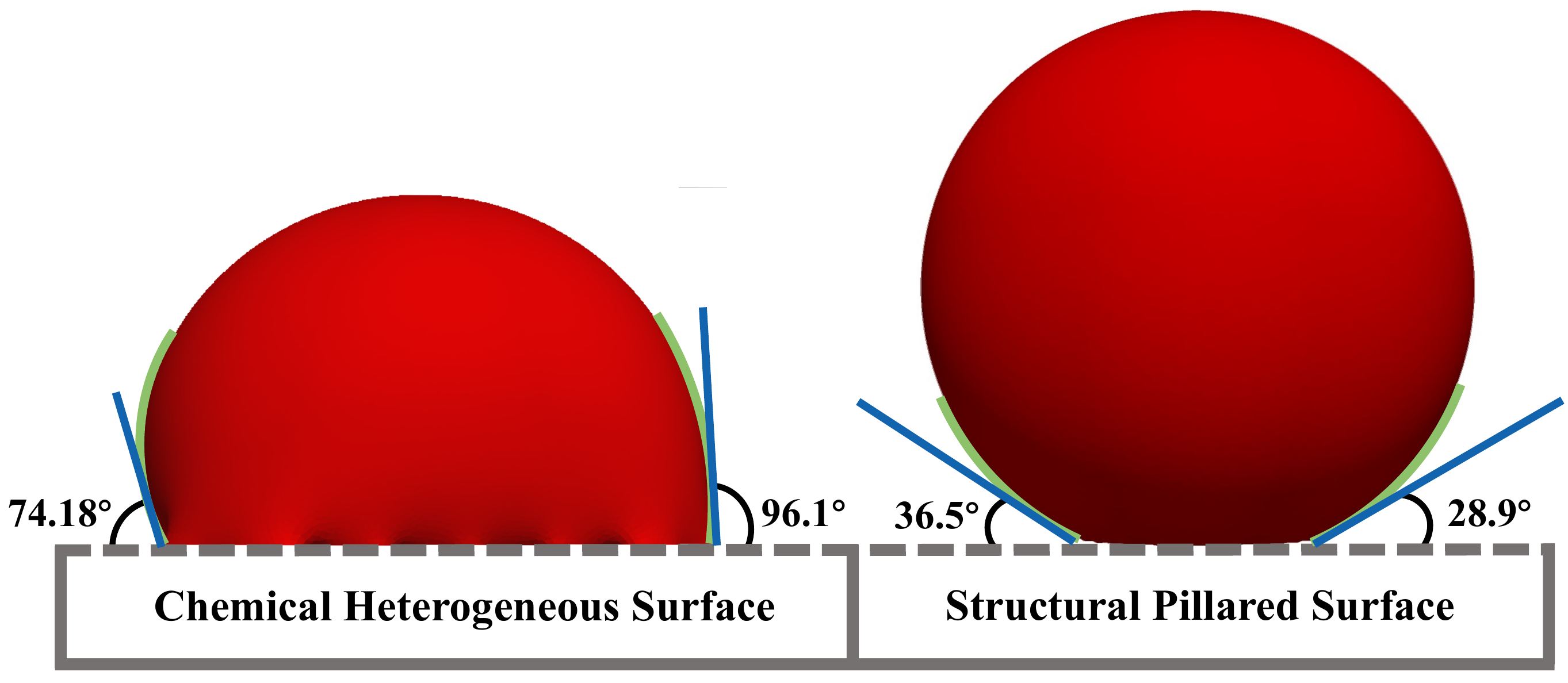}
\caption{Apparent contact angle measurements take from a random 2D plane of the simulation results. The green line represents the fit of the Young-Laplace equation to the fluid interface and the blue line is the tangent at the three-phase contact point.}
\label{2D}
\end{figure}

The apparent contact angle, $\theta_{app}$, determined by fitting the Young-Laplace equation to the entire spherical cap is provided in Table \ref{table1}. $\theta_{app}$ is the angle that could be measured experimentally by an observer. For the rough surface, $\theta_{app}$ provides a value near that of the Cassie-Baxter model, which represents approximately the minimum microscopic contact angle along the contact line. For the chemically varied surfaces, $\theta_{app}$ provides a value near the mean of the microscopic contact angles. For experimental studies, however, it is common to collect only a 2D projection of the sessile drop. Taking an random plane from the simulations, the $\theta_{app}$ measures are provided in Figure \ref{2D}. These measures obviously differ from that measured when the full 3D interface is considered, which demonstrates how experimentally measured contact angles can depend on the reference plane/cross section from which they are taken. 

\begin{table}
\centering
\caption{A comparison of effective contact angle measurements for the sessile drop simulations.}
\begin{tabular}{@{}p{2.0cm}p{1.65cm}p{1.65cm}p{1.65cm}@{}}
\toprule
\makecell[c]{Simulation} & $\theta_{CB}$ & $\theta_{app}$ & $\theta^{macro}$ \\ \midrule
Roughness & $28.1^{\circ}$ & $28.3^{\circ}$ & $34.5^{\circ}$\\
Chemical & $90.1^{\circ}$ & $79.2^{\circ}$ & $98.4^{\circ}$ \\ \bottomrule
\end{tabular}
\label{table1}
\end{table}
 
From the results, $\theta_{CB}$ deviates from the locally measured contact angles and is smaller than $\theta^{macro}$ for the tested surfaces. Furthermore, $\theta_{CB}$ only indicates the relationship of the apparent contact angle and fluid interfacial areas within the contact line loop, however, it does not capture any information on advancing and receding contact angles along the contact line. Recent studies have demonstrated that contact angles are determined by the linear fractions of solid and liquid along the contact line (even though it is extremely difficult to be predicted due to the contorted nature of contact line), not by the overall areal fractions of interfaces \cite{johnson1964contact,gao2007wenzel,extrand2002model,choi2009modified}. These issues arise when the length scale of the droplet and that of the surface features are not separable. Being a purely geometrical descriptor, $\theta^{macro}$ does not have this issue. Overall, the Cassie-Baxter relation predicts only a single value of an estimated contact angle, and consequently, this relation is inherently unable to provide an explanation for the observation of contact angle hysteresis, which indicates a limitation of the Cassie-Baxter model.

The macroscopic contact angle, $\theta^{macro}$, provides a macroscopic measure and captures the "average" effect of wetting hysteresis due to surface chemistry and geometry, as explained in previous works \cite{sun2020characterization,sun2020probing}. For both the chemically heterogeneous surface and structural rough surface, $\theta^{macro}$ captured a representative value of the local microscopic contact angles. In contrast, the Cassie-Baxter model had to be adjusted based on the type of surface structure rather than providing a single representative average of the local angles. In addition, depending on how the fluid droplets wets the surface other wetting models would need to be considered, i.e., the Wenzel model or wicking state model \cite{bico2002wetting}. Consequently, it can be elucidated that a universality of wetting state can be developed from the topological principles to account for wetting hysteresis originating from a multiscale surface regardless of the type of structure that is encountered and how the fluid wets that structure.

\section{Conclusions}
We hypothesised that a universal description of wetting on multiscale surfaces can be developed by using integral geometry coupled to thermodynamic laws \cite{de2013non,mecke2000additivity}. We demonstrated that the combination of surface energy minimization with the concept of deficit curvature from the Gauss-Bonnet theorem provides a universal description of wetting. The presented framework considers the link between fluid topology, curvature, and internal energy. It conceptually separates the different hierarchy levels of physical description from the thermodynamic aspects that can be added consistently via variational approaches. In principle, the presented concept is valid for any arbitrary geometry and reversible process, while the classic wetting models are all limiting cases of that universal picture.

A variational analysis was used to identify equilibrium conditions and demonstrate the universality of the developed framework. The minimum energy configuration was developed with no entropy production. While various technological applications are approximated well by reversible processes \cite{blunt2017multiphase}, other applications require the consideration of irreversible processes \cite{kjelstrup2017non,seth2006efficiency}. The variational analysis also considers a fluid droplet of constant topology with $\chi = 1$. For multiphase porous systems, a wide range of fluid typologies are possible \cite{schluter2016pore,armstrong2016beyond}. The limiting case of equilibrium and $\chi = 1$ were considered to demonstrate that the classic wetting models are obtained from the proposed framework. The development of wetting models for other situations is possible using the presented general framework of Section 3.

For multiphase systems, $\theta^{macro}$, as presented in Eq. (\ref{normal_single}), can be used to describe the wetting state. The wetting metric defines the amount of deficit curvature per contact line loop, and thus provides a macroscale constant for any wetting system. The metric was demonstrated by the simulation results presented in Figures \ref{chem} and \ref{rough} where the hysteresis loop of local microscopic contact angles varied above and below $\theta^{macro}$. The universality of $\theta^{macro}$ is that its usage is not contingent on any wetting state whereas traditional wetting models depend on how the given fluid wets the surface \cite{bico2002wetting}. Alternatively, the experimentally observed apparent contact angle depends on the reference plane of the measurement \cite{andrew2014pore}. Being of topological origin $\theta^{macro}$ would be applicable to describe any type of wetting phenomena in complex geometries. 

Understanding wetting behaviour provides a means to design heterogeneous surfaces for energy and microfluidic applications where the movement of liquid is controlled by capillary forces. Likewise, surface heterogeneity is common in geological systems where the movement of ground water and/or hydrocarbon is capillary controlled \cite{blunt2017multiphase}. These applications occur in multiscale hierarchical structures, which in addition to having multiscale surfaces are also topologically complex. The presented framework provides a way forward when dealing with the wettability of such systems. 

\section*{Acknowledgments}
C. S. acknowledges the Fundamental Research Funds for the Central Universities (No. 2462021BJRC004). R. A. and P. M. acknowledge DP210102689 Discovery grant from Australian Research Council. 
\bibliographystyle{elsarticle-num}

\bibliography{cas-refs}


\end{document}